\documentclass[10pt,conference]{IEEEtran}
\usepackage{ifpdf}
\usepackage{url}
\ifpdf
\else
\fi
\usepackage{cite}
\usepackage{balance}
\usepackage{enumitem}
\usepackage{bm,comment,color}

\ifCLASSINFOpdf
  \usepackage[pdftex]{graphicx}
  \graphicspath{{../pdf/}{../jpeg/}}
  \DeclareGraphicsExtensions{.pdf,.jpeg,.png}
\else
  \usepackage[dvips]{graphicx}
  \graphicspath{{../eps/}}
  \DeclareGraphicsExtensions{.eps}
\fi
\usepackage{amsmath}
\usepackage{algpseudocode}
\algtext*{EndWhile}
\algtext*{EndIf}
\algtext*{EndFor}
\usepackage{algorithm}
\usepackage{array}
\usepackage{amsfonts} 
\usepackage{amssymb}
\usepackage{esint} 
\usepackage{units}
\usepackage{multirow}
\usepackage{comment}

\usepackage{color}

\usepackage{standalone}

\usepackage{pgfplots}
\usepackage{tikz}
\usetikzlibrary{calc}
\makeatletter
\newcommand{\gettikzxy}[3]{%
  \tikz@scan@one@point\pgfutil@firstofone#1\relax
  \edef#2{\the\pgf@x}%
  \edef#3{\the\pgf@y}%
}
\usetikzlibrary{spy,backgrounds}
\pgfplotsset{compat=newest}
\usetikzlibrary{plotmarks}
\usetikzlibrary{arrows.meta}
\usepgfplotslibrary{patchplots}
\usepackage{grffile}
\newlength\fheight 
\newlength\fwidth 
\usepgfplotslibrary{fillbetween}

\usepackage{acronym}

\acrodef{6g}[6G]{the sixth generation}
\acrodef{aoa}[AOA]{angle-of-arrival}
\acrodef{bs}[BS]{base station}
\acrodef{bse}[BSE]{beam squint effect}
\acrodef{crb}[CRB]{Cram\'er-Rao bound}
\acrodef{elaa}[ELAA]{extremely large antenna array}
\acrodef{ff}[FF]{far-field}
\acrodef{las}[L\&S]{localization and sensing}
\acrodef{los}[LOS]{line-of-sight}
\acrodef{nf}[NF]{near-field}
\acrodef{nlos}[NLOS]{non-line-of-sight}
\acrodef{ofdm}[OFDM]{orthogonal frequency division multiplexing}
\acrodef{ris}[RIS]{reconfigurable intelligent surface}
\acrodef{sns}[SNS]{spatial non-stationarity}
\acrodef{swm}[SWM]{spherical wave model}
\acrodef{siso}[SISO]{single-input single-output}
\acrodef{ue}[UE]{user equipment}
\acrodef{dmimo}[D-MIMO]{distributed MIMO}

\usepackage{tabu,longtable}

\ifCLASSOPTIONcompsoc
 \usepackage[caption=false,font=normalsize,labelfont=sf,textfont=sf]{subfig}
\else
 \usepackage[caption=false,font=footnotesize]{subfig}
\fi
\usepackage{stfloats}
\usepackage[bookmarks=false, hidelinks]{hyperref}
\usepackage{xcolor}
\long\def\comment#1{}

\newfont{\bbb}{msbm10 scaled 700}

\newfont{\bb}{msbm10 scaled 1100}

\usepackage{graphicx}      

\usepackage{geometry}
\renewcommand{\baselinestretch}{0.98}
\begin{document}

\bstctlcite{IEEEexample:BSTcontrol}

\title{
Observation Compression in Rate-Limited Closed-Loop Distributed ISAC Systems: From Signal Reconstruction to Control
}
\IEEEoverridecommandlockouts
\author{
Guangjin Pan\textsuperscript{*}, Zhixing Li\textsuperscript{†}, Ayça Özçelikkale\textsuperscript{‡}, Christian Häger\textsuperscript{*}, Musa Furkan Keskin\textsuperscript{*}, Henk Wymeersch\textsuperscript{*}\\
\textsuperscript{*}Department of Electrical Engineering, Chalmers University of Technology, Sweden \\
\textsuperscript{†}Department of Computer Science and Engineering, Chalmers University of Technology, Sweden \\
\textsuperscript{‡}Department of Electrical Engineering, Uppsala University, Sweden
\vspace{-8mm}
\thanks{This work was supported in part by a grant from the Chalmers AI Research
 Center Consortium (CHAIR), by the National Academic Infrastructure for
 Supercomputing in Sweden (NAISS), by the SNS JU project 6G-DISAC under the EU's Horizon Europe research and innovation Program under Grant Agreement No 101139130,  the Swedish Foundation for Strategic Research (SSF) (grant FUS21-0004, SAICOM), the Swedish Research Council (VR) through the project 6G-PERCEF under Grant 2024-04390, and the Chalmers Areas of Advance in ICT and Transport.}}

\maketitle

\thispagestyle{empty}


\begin{abstract}
In closed-loop distributed multi-sensor integrated sensing and communication (ISAC) systems, performance often hinges on transmitting high-dimensional sensor observations over rate-limited networks. In this paper, we first present a general framework for rate-limited closed-loop distributed ISAC systems, and then propose an autoencoder-based observation compression method to overcome the constraints imposed by limited transmission capacity. Building on this framework, we conduct a case study using a closed-loop linear quadratic regulator (LQR) system to analyze how the interplay among observation, compression, and state dimensions affects reconstruction accuracy, state estimation error, and control performance. In multi-sensor scenarios, our results further show that optimal resource allocation initially prioritizes low-noise sensors until the compression becomes lossless, after which resources are reallocated to high-noise sensors.

\end{abstract}

\begin{IEEEkeywords}
Distributed ISAC system, autoencoder, 5G/6G networks, closed-loop system.
\end{IEEEkeywords}

\IEEEpeerreviewmaketitle
\acresetall 
\vspace{0mm}
\section{Introduction}

Recent advances in 5G/6G networks have enabled closed-loop distributed integrated sensing and communication (ISAC) systems \cite{Strinati_DISAC_2025}, where sensing, communication, and control are tightly coupled in real time \cite{Park_WNCS_2018}. These systems show strong potential in applications such as autonomous driving \cite{Wang_AutonomousDriving_2019}, industrial automation \cite{Liang_Factory_Automation_2019}, and UAVs \cite{Chang_IntegratedUAV_2022}. Unlike conventional decoupled designs, closed-loop ISAC jointly optimizes sensing, communication, and control for robust estimation and decision-making under rate limits, enhancing efficiency and resilience amid bandwidth and network dynamics..

In such systems, a key challenge lies in the degradation of sensing quality caused by rate-limited links, which in turn affects control performance. To address this, model-based methods such as principal component analysis (PCA) \cite{Nagashima_PCA_2016} and AI-based approaches like autoencoders (AEs) \cite{Ismayilov_AE_2024} reduce the dimensionality of signals (e.g., radar or channel observations). In parallel, emerging research on semantic \cite{Uysal_Semantic_2022, girgis_semantic_2024} and goal-oriented communication \cite{Wu_Goal-Oriented_2024, Wang_Goal-Oriented_2024} focuses on transmitting only task-relevant information, rather than preserving full signal fidelity. While prior work has addressed observation compression in single-sensor \cite{Wang_Goal-Oriented_2024, girgis_predictive_2022, Sabag_Reducing_2023} and multi-sensor \cite{Rego_Distributed_2021} closed-loop ISAC system, relatively few studies have systematically investigated the interplay among observation dimension, compression strategy, and control performance in general closed-loop ISAC systems under rate-limited multi-sensor settings.

In this work, we aim to bridge this gap by developing and analyzing a general framework for rate-limited closed-loop distributed ISAC systems. Our contributions are: (i) We propose a general closed-loop distributed ISAC framework integrating sensing, compression, state estimation, and control, with detailed descriptions of each component. (ii) We formulate the observation compression problem in closed-loop distributed ISAC systems under rate-limited communication, and present an AE-based method for learning compact representations. (iii) We conduct a detailed case study based on a linear quadratic regulator (LQR) control problem to investigate the interplay between state, observation, and compression dimensions, and further explore optimal transmission resource allocation strategies in multi-sensor scenarios.

\begin{figure}[tb]
\centering
 \includegraphics[scale=0.32]{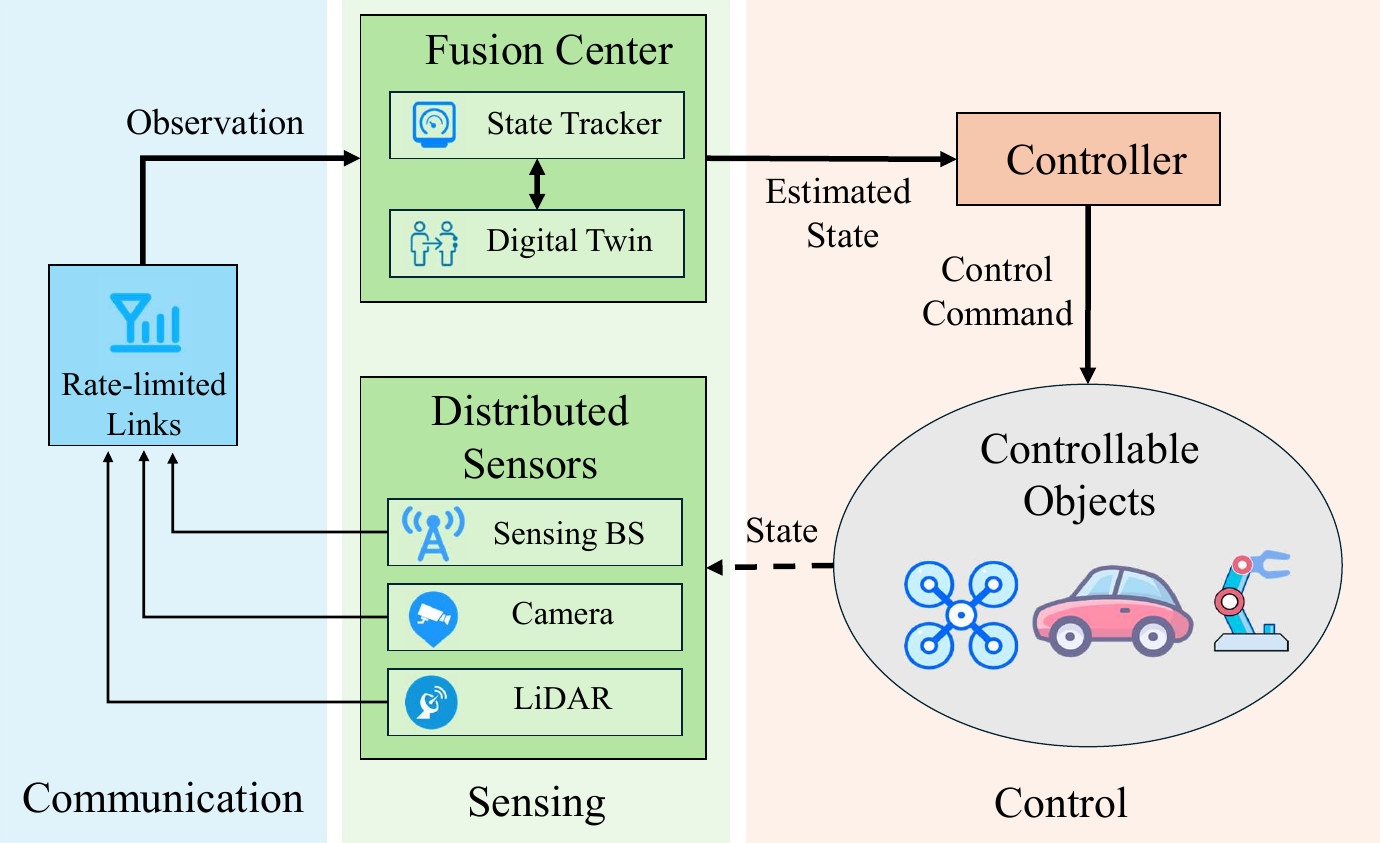}
 \vspace{-1mm}
\caption{System architecture of a closed-loop distributed ISAC system with rate-limited links.}
\label{fig:DISACsystem}
\vspace{-5mm}
\end{figure}

\vspace{-2mm}
\section{Closed-loop Distributed ISAC System Model}

Closed-loop distributed ISAC systems, as depicted in Fig.~\ref{fig:DISACsystem}, integrate sensing, communication, and control, with applications in autonomous driving and UAV coordination. This section outlines a general framework for such systems.
\begin{itemize}[leftmargin=*, topsep=0pt, itemsep=0pt, parsep=0pt, partopsep=0pt]
    \item 

\textit{State Model:}
We consider an environment with controllable objects, such as cars. The combined state of the controllable objects is denoted by $\bm{x}_{t}$ and is governed by a first-order Markov model
$\bm{x}_{t+1}\sim p_{\text{obj}}(\bm{x}_{t+1}|\bm{x}_{t},\bm{u}_{t})$,
where $\bm{u}_{t}$ is a control command and $p_{\text{obj}}(\bm{x}_{t+1}|\bm{x}_{t},\bm{u}_{t})$
represents the object dynamics and response to the control command. This control command could be in the form of steering and acceleration. The true state is not directly observable by the network but can be inferred via sensors.

\item
\textit{Sensor Model:}
The environment is equipped with $S$ sensors, (e.g., full-duplex ISAC-enabled base stations \cite{he2023full}) that continuously sense and track the controllable objects. These sensors collect high-dimensional raw observations (such as channel measurements, images, or dense point clouds) and forward them to a fusion center for state estimation and prediction. We denote the observation from sensor $s\in\{1,\ldots,S\}$ by 
$\bm{y}_{t}^{s}\sim p_{\text{meas}}^{s}(\bm{y}_{t}^{s}|\bm{x}_{t})$,
where $p_{\text{meas}}^{s}(\cdot|\bm{x}_{t})$ is the observation model
of sensor $s$.
\item
\textit{Digital Twin and State Tracker:}
The fusion center maintains a real-time model of the tracked objects
in the form of a digital twin (DT). Given observations $\mathcal{Y}_{t}=\bm{y}_{t}^{1:S}$ and controls $\mathcal{U}=\bm{u}_{1:t}$, the DT models the posterior 
$q(\bm{x}_{t}|\mathcal{Y}_t,\mathcal{U}_{t-1})$,
and the kinematics
$p_{\text{DT}}(\bm{x}_{t+k}|\bm{x}_{t},\bm{u}_{t})$,
for any $k>0$. Note that $p_{\text{DT}}(\bm{x}_{t+k}|\bm{x}_{t},\bm{u}_{t})$
may or may not be identical to $p_{\text{obj}}(\bm{x}_{t+k}|\bm{x}_{t},\bm{u}_{t})$,
depending on the fidelity of the DT. The DT then provides a prediction
on the states of the tracked objects. Given observations $\mathcal{Y}{t+1}$ with measurement model
$p_{\text{DT}}(\mathcal{Y}_{t+1}|\bm{x}_{t+1})=\prod_{s=1}^{S}p_{\text{meas}}^{s}(\bm{y}_{t+1}^{s}|\bm{x}_{t+1})$,
the state tracker updates the posterior as \vspace{-2mm}
\begin{align}
q(\bm{x}_{t+1}|\mathcal{Y}_{t+1},\mathcal{U}_{t})\propto  & \ p_{\text{DT}}(\mathcal{Y}_{t+1}|\bm{x}_{t+1})  \\ 
& \! \! \! \! \! \! \!  \! \int \! \! q(\bm{x}_{t}|\mathcal{Y}_t,\mathcal{U}_{t-1})p_{\text{DT}}(\bm{x}_{t+1}|\bm{x}_{t},\bm{u}_{t})\mathrm{d}\bm{x}_{t}.\label{eq:dynamics}\nonumber
\end{align} \vspace{0mm}
Deploying a DT enhances system performance by evaluating the impact of measurement distortion, aggregating environmental information, forecasting dynamics, and
allocating transmission resources among multiple sensors to maintain
accurate state estimation \cite{Chen_DigitalTwin_2024, Khan_DigitalTwin_2022}.
\item
\textit{Controller:}
Assuming a separation between state tracking and control, the controller
takes as input the posterior distribution from the DT and
determines the control command: 
$\bm{u}_{t+1}\sim\pi(q(\bm{x}_{t+1}|\mathcal{Y}_{t+1},\mathcal{U}_{t}))$,
where $\pi(\cdot)$ denotes the control policy. This policy is determined to optimize a control objective, using methods such as model predictive control (MPC) or reinforcement learning (RL). The resulting commands pursue goals like guiding vehicles to target positions or stabilizing UAV trajectories. These commands are then transmitted to the controlled objects, influencing subsequent state transitions via $p_{\text{obj}}(\bm{x}{t+2}|\bm{x}{t+1},\bm{u}_{t+1})$.
\item
\textit{Rate-Limited Links:}
Given the significant communication constraints typical in distributed systems, directly transmitting these raw observations $\bm{y}_{t}^{1:S}$ could rapidly exceed the available network bandwidth \cite{Strinati_DISAC_2025, Guo_Toward_2025, pan2025ai}. Therefore, sensors must apply \emph{local data compression methods} and resource allocation schemes, extracting and transmitting only task-relevant information, thus maintaining accurate state estimation at the fusion center without violating bandwidth limitations. Goal-oriented semantic communication strategies have emerged as promising solutions to enhance data compression efficiency by capturing the most relevant information directly related to the control objectives.
\end{itemize}

\vspace{-1mm}
\section{Autoencoder-Based Compression Framework}

In this section, we formalize the compression problem in rate-limited closed-loop distributed ISAC systems and propose an AE-based framework for learning compact representations.

\vspace{-1mm}

\subsection{Rate-Limited Problem Description}

As bandwidth is limited in closed-loop distributed ISAC systems, sensors locally compress their observations and collaboratively allocate resources to ensure that critical information reaches the fusion center. Specifically, each sensor $s$ encodes its raw observation $\bm{y}_t^s$ into a compressed representation $\bm{z}_t^s = \mathcal{F}_{\text{comp}}^s(\bm{y}_t^s; \bm{\theta}_{\text{comp}}^s)$ using a local compression function $\mathcal{F}_{\text{comp}}^s$, parameterized by $\bm{\theta}_{\text{comp}}^s$. The compressed data is transmitted over a rate-limited link\footnote{For the sake of clarity, these links are assumed to be error-free and do not incur any latency.} to the fusion center, where it is reconstructed via a corresponding decompression function $\mathcal{F}_{\text{rec}}^s$ with parameters $\bm{\theta}_{\text{rec}}^s$. The reconstruction process is defined as $\hat{\bm{y}}_t^s = \mathcal{F}_{\text{rec}}^s(\bm{z}_t^s; \bm{\theta}_{\text{rec}}^s)$.
To meet communication constraints, the total transmission dimension from all sensors at each time step must satisfy
    $\sum_{s=1}^{S} \dim(\bm{z}_t^s) \le r_t,$
where $\dim(\bm{z}_t^s)$ denotes the \textit{compression dimension} for sensor $s$. Generally, both the transmission dimension budget $r_t$ and the compression dimension $\bm{z}_t^s$ may vary over time, enabling dynamic adaptation to channel conditions, task requirements, or estimation uncertainty.\footnote{While compressed observations reduce transmission delay and thereby reduce sensing latency, the effects of control frequency and sensing latency are not considered in this work \cite{moayedi2009adaptive}.}

\vspace{-1mm}
\subsection{Three Levels of Error in Distributed ISAC} \label{sec:error-levels}

To characterize performance under rate-limited communication, we consider three error levels, each tied to a stage of the closed-loop ISAC pipeline:
\begin{itemize}
    \item \textit{L1 -- Reconstruction Error:} Measures the difference between the original and reconstructed observations, indicating the compression and reconstruction effectiveness.
    \item \textit{L2 -- Estimation Error:} Quantifies the accuracy of the state estimation by measuring the deviation between the fusion center’s estimation and the true system state, reflecting the impact of compression fidelity on tracking performance.
    \item \textit{L3 -- Control Error:} Captures the ultimate impact of compression on system objectives by evaluating the incurred control cost within the closed-loop operation, directly linking compression quality to control effectiveness.
\end{itemize}
This multi-level error framework provides a comprehensive perspective for analyzing how different compression strategies affect reconstruction fidelity, state estimation accuracy, and ultimately, closed-loop control performance.

\vspace{-1mm}

\subsection{Autoencoder-Based Compression Method}

In this study, we focus on designing compression methods that explicitly optimize for L1 -- reconstruction error.\footnote{Although distributed sensors may cause interference or complementary gains, this work focuses on observation compression in closed-loop distributed ISAC systems and ignores such effects. Future work will incorporate task-driven losses tied to estimation or control objectives.}  We focus on how minimizing reconstruction error impacts state estimation and control performance in closed-loop distributed ISAC systems. To this end, we develop an AE-based compression framework tailored for minimizing reconstruction error. 
An AE is a neural network architecture composed of an encoder and a decoder \cite{li2023comprehensive}. The encoder compresses the input observation into a low-dimensional latent representation. The decoder attempts to reconstruct the original observation from this compressed data.

For each sensor, given a fixed compression dimension, the AE is trained to minimize the long-term mean squared error (MSE) between the original observation and its reconstruction:
\begin{align} 
\! \! \mathcal{L}_{\text{MSE}}^s \! = \! \lim_{T \to \infty} \!  \frac{1}{T} \! \sum_{t=1}^{T} \left| \bm{y}_t^s  \! - \! \mathcal{F}_{\text{rec}}^s \left( \mathcal{F}_{\text{comp}}^s(\bm{y}_t^s; \bm{\theta}_{\text{comp}}^s); \bm{\theta}_{\text{rec}}^s \right) \right|^2 \! \! .
\end{align}
The AE learns a compact latent space tailored to the data distribution under rate-limited constraints.

\vspace{-1mm}
\section{Case Study: Closed-Loop LQR}

\begin{figure}[tb]
    \centering
    \includegraphics[width=0.9\linewidth]{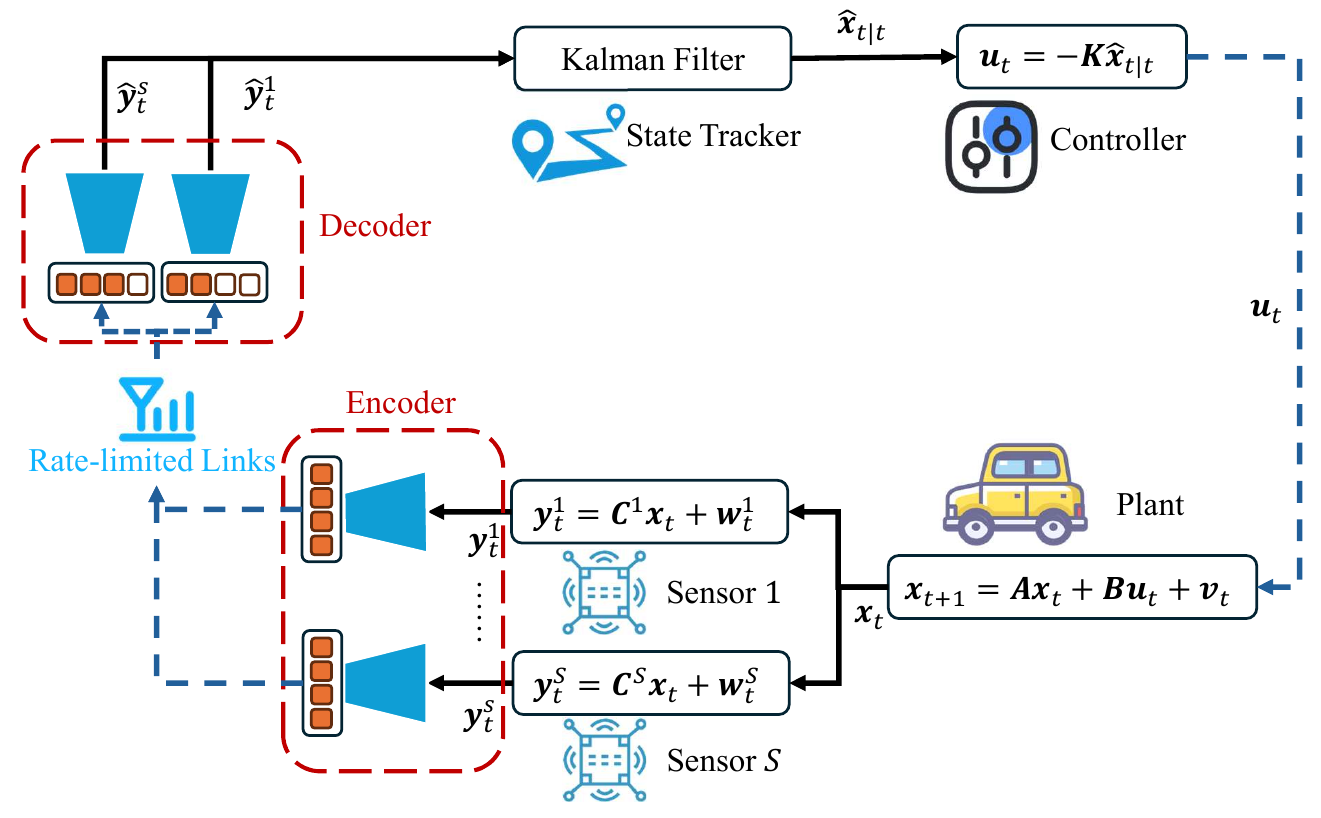}
    \vspace{-2mm}
    \caption{Closed-loop Multi-sensor LQR system with rate-limited links.}
    \label{fig_LQR}
    \vspace{-5mm}
\end{figure}

To evaluate the proposed compression framework, we study a closed-loop LQR problem \cite{Sabag_Reducing_2023}, a fundamental paradigm in control theory that provides closed-form solutions and strong theoretical guarantees, and serves as a representative instance of the distributed ISAC setting. It serves as a first-order approximation for many more complex nonlinear control problems and has been extensively used to study estimation and control under communication constraints~\cite{Sabag_Reducing_2023}. The schematic diagram of the LQR system is shown in Figure~\ref{fig_LQR}. Our goals are: (i) to analyze how system performance depends on the interplay among state, observation, and compression dimensions; and (ii) to study how to allocate limited communication resources across sensors with different noise levels to enhance control performance.

\vspace{-1mm}
\subsection{Scenario Description}

In distributed LQR systems, sensors send rate-limited observations of the plant to a fusion center, where a Kalman filter estimates the state. The controller then generates commands from this estimate, forming the closed loop.

\subsubsection{State Model}
The system dynamics are given by $
    \bm{x}_{t+1} = \bm{A} \bm{x}_{t} + \bm{B} \bm{u}_{t} + \bm{v}_{t}$,
where $\bm{x}_{t} \in \mathbb{R}^{N_x}$ is the state of dimension $N_x$, and $\bm{u}_t \in \mathbb{R}^{N_u}$ is the control of dimension $N_u$. The matrices $\bm{A} \in \mathbb{R}^{N_x \times N_x}$ and $\bm{B} \in \mathbb{R}^{N_x \times N_u}$ represent the known system dynamics and control input mappings, respectively. The process noise $\bm{v}_t \sim \mathcal{N}(0, \bm{Q})$ is with covariance matrix $\bm{Q} \in \mathbb{R}^{N_x \times N_x}$. At each time step, sensor $s$ provides linear measurements:
\begin{align}
    \bm{y}_{t}^s = \bm{C}^s \bm{x}_{t} + \bm{w}_{t}^s,
    \label{equ:observation}
\end{align}
where $\bm{y}_t^s \in \mathbb{R}^{N^s_y}$ is the observation vector from sensor $s$ with observation dimension $N^s_y$, and $\bm{C}^s \in \mathbb{R}^{N^s_y \times N_x}$ is the known observation matrix. The observation noise $\bm{w}_t^s \sim \mathcal{N}(0, \bm{R}^s)$ is with covariance $\bm{R}^s \in \mathbb{R}^{N^s_y \times N^s_y}$. The compressed data $\bm{z}_t^s$ is obtained using $\bm{z}_t^s = \mathcal{F}_{\text{comp}}^s(\bm{y}_t^s; \bm{\theta}_{\text{comp}}^s)$.

\subsubsection{State Tracker}
Prior to the state tracker, the compressed measurements are decompressed using $\hat{\bm{y}}_t^s = \mathcal{F}_{\text{rec}}^s(\bm{z}_t^s; \bm{\theta}_{\text{rec}}^s)$, and then provided to the Kalman filter. The Kalman filter combines them with the system model to estimate $\hat{\bm{x}}_{t|t}$ and its associated covariance matrix $\bm{\Sigma}_{t|t}$, following prediction and update steps \cite{khodarahmi2023review}. The prediction step relies on the DT of the real system \eqref{eq:dynamics}. 
When the observation $\bm{y}_t^s$ is perfectly reconstructed and the DT models the system correctly, the Kalman filter is optimal. However, when there is reconstruction error in $\hat{\bm{y}}_t^s$, the Kalman filter becomes suboptimal.

\subsubsection{Controller}
In the controller, the LQR computes the control input $\bm{u}_{t}$ by minimizing the infinite-horizon cost:
\begin{equation} 
    J = \sum_{t=1}^\infty \Bigl( \tilde{\bm{x}}_t^\top \bm{Q}_{\mathrm{goal}}  \tilde{\bm{x}}_t 
    + \bm{u}_t^\top \bm{R}_{\mathrm{goal}} \bm{u}_t \Bigr),
    \label{eq:lqrcost}
\end{equation}
where $\tilde{\bm{x}}_t = \bm{x}_t - \bm{x}_{\mathrm{desired}}$ denotes the deviation from the target state, and $\bm{Q}_{\mathrm{goal}}$ and $\bm{R}_{\mathrm{goal}}$ are matrices penalizing state deviation and control effort, respectively. When the observation $\bm{y}$ is perfect, the optimal control input follows the LQR policy: $\bm{u}_{t} = -\bm{K}\,\hat{\bm{x}}_{t|t}$,
where $\bm{K}$ can be computed as described in \cite{shaiju2008formulas}.

Without compression, the LQR system achieves optimal performance. However, compression over rate-limited links introduces errors that may destabilize the system. In our experiments, we fix the LQR parameters so that the system remains stable with a stationary data distribution when the reconstruction error is below a critical threshold. This setup enables us to focus on the static impact of compression dimensions and resource allocation strategies on performance, without dynamic variations.

\vspace{-1mm}
\subsection{Experimental Setup and Baselines}

\subsubsection{LQR Parameter}We consider a vehicle control scenario in our experiments. For the plant, we set 
\begin{align}
    \bm{A}=\begin{bmatrix}
  1 & 0 & \Delta t & 0\\
  0 & 1 & 0 & \Delta t\\
  0 & 0 & 1 & 0\\
  0 & 0 & 0 & 1  
  \end{bmatrix},
  \bm{B}=\begin{bmatrix}
  0.5 \Delta t^2  & 0  \\
  0 & 0.5 \Delta t^2 \\
  \Delta t & 0  \\
  0 & \Delta t  
  \end{bmatrix}, \nonumber
\end{align}
with $\Delta t=0.1$, $N_x$ = 4. The state $\bm{x}_t\in\mathbb{R}^4$ represents 2D position and velocity, and the control $\bm{u}_t\in\mathbb{R}^2$ represents the acceleration. Each sensor observes $\bm{x}_t$ via 
\eqref{equ:observation} where $\bm{C}^s$ is generated once and fixed, with each element independently drawn from $\mathcal{N}(0, {1}/{50})$. The observation noise is set as $\bm{w}_t^s \sim \mathcal{N}(0,\bm{I})$. We also set $\hat{\bm{x}}_{0|0} = \bm{x}_0$, $\bm{\Sigma}_{0|0} = 0.0001\,\bm{I}$, $\bm{Q} = \bm{I}$, and $\bm{R}^s = \bm{I}$. The LQR cost uses $\bm{Q}_{\mathrm{goal}} = 0.1\,\bm{I}$ and $\bm{R}_{\mathrm{goal}} = \bm{I}$.

\subsubsection{AE Implementation}In our implementation, both the encoder and decoder are designed as fully-connected multi-layer perceptrons. The encoder consists of four linear layers with ReLU activations, with the number of neurons in each layer set to 512, 1024, 512, and the target compression dimension $\dim(\bm{z}_t^s)$, respectively. The decoder mirrors this structure symmetrically to reconstruct the observation from the latent representation. We run 1000 rounds of 200 steps each to collect 200{,}000 training samples, then test on another 1000 rounds (200 steps each). We compare the AE with classical PCA, expected to be optimal for this linear problem, aiming to show that the AE can at least match PCA’s performance.

\subsubsection{Evaluated Methods}To simplify the descriptions in experiments, we define the following testing mechanisms.
\begin{itemize}
    \item \textbf{AE/PCA-online:} The AE/PCA algorithm compresses and decompresses in a closed-loop system, where poor compression performance can lead to accumulated error.
    \item \textbf{AE/PCA-offline:} The AE/PCA algorithm is applied to previously collected data rather than used in closed-loop experiments, it does not affect closed-loop performance.
    \item \textbf{UC:}
Denotes the uncompressed case where the observation dimension equals the x-axis compression dimension.
    \item \textbf{UC (o):}
Denotes the uncompressed case with the observation dimension fixed at value o.
\end{itemize}

In line with the three-level compression framework from Section \ref{sec:error-levels}, we consider three levels of error: reconstruction error, measured by $|\hat{\bm{y}}_t^s - \bm{y}_t^s|^2$; state estimation error, given by $|\hat{\bm{x}}_{t|t} - \bm{x}_t|^2$; and the LQR cost, as defined in \eqref{eq:lqrcost}.
\color{black}

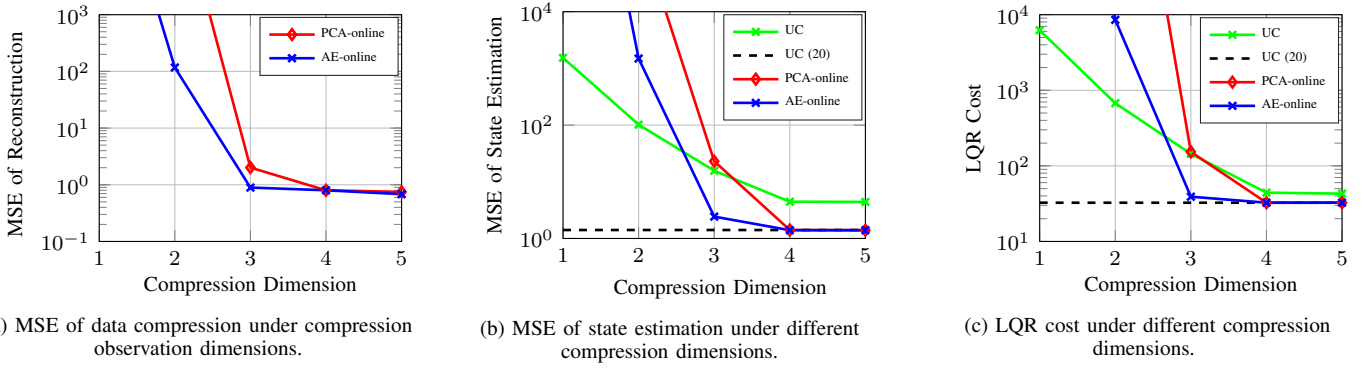
\begin{figure*}[t]
    \centering
    \captionsetup[subfigure]{justification=centering, font=footnotesize, skip=5pt} 
    \begin{minipage}{0.32\textwidth}        
    	\subfloat[MSE of data compression under compression observation dimensions.\label{fig:MSE_ObsDim20_diffComp}]{
%
\newcommand\ms{0.8}
\newcommand\lw{1.0}
\definecolor{mycolor1}{rgb}{0.00000,1.00000,1.00000}%
\begin{tikzpicture}[font=\footnotesize, spy using outlines={rectangle, magnification=2.5, size=0.3cm, connect spies}]
\begin{axis}[%
width=4.cm,
height=3.cm,
at={(0in,0in)},
scale only axis,
xmin=1,
xmax=5,
xlabel style={yshift=0.5 ex},
xlabel={Compression Dimension},
ymode=log,
ymin=0.1,
ymax=1e3,
ytick={1e-1, 1e0, 1e1, 1e2, 1e3},
ylabel style={xshift=-0.5 ex},
ylabel={MSE of Reconstruction},
axis background/.style={fill=white},
xmajorgrids,
ymajorgrids,
legend columns=1, 
legend style={font=\tiny, at={(0.53, 0.73)}, anchor=south west, legend cell align=left, align=left, draw=white!15!black}
]

\addplot [color=red, line width=\lw pt, mark=diamond, mark options={solid, red}]
  table[row sep=crcr]{%
1  6e8 \\
2  2e5  \\
3  2.01 \\
4  0.799 \\
5  0.749 \\
};
\addlegendentry{PCA-online}

\addplot [color=blue, line width=\lw pt, mark=x, mark options={solid, blue}]
  table[row sep=crcr]{%
    1  3e6 \\
    2  117  \\
    3  0.892 \\
    4  0.800 \\
    5  0.682 \\
};
\addlegendentry{AE-online}

\end{axis}

\end{tikzpicture}%
}\\
    \end{minipage}
    \hfill
    \begin{minipage}{0.32\textwidth}    
    	\subfloat[MSE of state estimation under different compression
    dimensions.\label{fig:state_ObsDim20_diffComp}]{
%
\newcommand\ms{0.8}
\newcommand\lw{1.0}
\definecolor{mycolor1}{rgb}{0.00000,1.00000,1.00000}%
\begin{tikzpicture}[font=\footnotesize, spy using outlines={rectangle, magnification=2.5, size=0.3cm, connect spies}]
\begin{axis}[%
width=4.cm,
height=3.cm,
at={(0in,0in)},
scale only axis,
xmin=1,
xmax=5,
xlabel style={yshift=-0.0 ex},
xlabel={Compression Dimension},
ymode=log,
ymin=1,
ymax=1e4,
xlabel style={xshift=0.5 ex},
ylabel={MSE of State Estimation},
axis background/.style={fill=white},
xmajorgrids,
ymajorgrids,
legend columns=1, 
legend style={font=\tiny, at={(0.53, 0.52)}, anchor=south west, legend cell align=left, align=left, draw=white!15!black}
]
\addplot [color=green, line width=\lw pt, mark=x, mark options={solid, green}]
  table[row sep=crcr]{%
1 1537  \\
2 101.8 \\
3 15.66 \\
4 4.450 \\
5 4.412  \\
6 3.101 \\
};
\addlegendentry{UC}

\addplot [color=black, line width=\lw pt, mark=., mark options={solid, black}, dashed]
  table[row sep=crcr]{%
1 1.41 \\
6 1.41 \\
};
\addlegendentry{UC (20)}

\addplot [color=red, line width=\lw pt, mark=diamond, mark options={solid, red}]
  table[row sep=crcr]{%
1  7e7 \\
2  3e5  \\
3  23.0 \\
4  1.41 \\
5  1.40 \\
6  1.40 \\
};
\addlegendentry{PCA-online}

\addplot [color=blue, line width=\lw pt, mark=x, mark options={solid, blue}]
  table[row sep=crcr]{%
  1 8e9 \\
  2 1493 \\
  3 2.43 \\
  4 1.40 \\
  5 1.40\\ 
  6 1.40 \\
};
\addlegendentry{AE-online}

\end{axis}

\end{tikzpicture}%
}\\
    \end{minipage}
    \hfill
    \begin{minipage}{0.32\textwidth}
    	\subfloat[LQR cost under different compression dimensions. \label{fig:LQR_ObsDim20_diffComp}]{
%
\newcommand\ms{0.8}
\newcommand\lw{1.0}
\definecolor{mycolor1}{rgb}{0.00000,1.00000,1.00000}%
\begin{tikzpicture}[font=\footnotesize, spy using outlines={rectangle, magnification=2.5, size=0.3cm, connect spies}]
\begin{axis}[%
width=4.cm,
height=3.cm,
at={(0in,0in)},
scale only axis,
xmin=1,
xmax=5,
xlabel style={yshift=0.5 ex},
xlabel={Compression Dimension},
ymode=log,
ymin=10,
ymax=1e4,
ylabel style={yshift=-0.5 ex},
ylabel={LQR Cost},
axis background/.style={fill=white},
xmajorgrids,
ymajorgrids,
legend columns=1, 
legend style={font=\tiny, at={(0.53, 0.52)}, anchor=south west, legend cell align=left, align=left, draw=white!15!black}
]
\addplot [color=green, line width=\lw pt, mark=x, mark options={solid, green}]
  table[row sep=crcr]{%
1 6221.5  \\
2 675.60\\
3 144.30\\
4 44.20\\
5 42.92\\
6 40.70 \\
};
\addlegendentry{UC}

\addplot [color=black, line width=\lw pt, mark=., mark options={solid, black},dashed]
  table[row sep=crcr]{%
1 32.56 \\
6 32.56 \\
};
\addlegendentry{UC (20)}

\addplot [color=red, line width=\lw pt, mark=diamond, mark options={solid, red}]
  table[row sep=crcr]{%
1  3e9 \\
2  1e8  \\
3  153.27 \\
4  32.56 \\
5  32.56 \\
6  32.56 \\
};
\addlegendentry{PCA-online}

\addplot [color=blue, line width=\lw pt, mark=x, mark options={solid, blue}]
  table[row sep=crcr]{%
1  3e9 \\
2  8547 \\
3  39.1 \\
4  32.57 \\
5  32.60 \\
};
\addlegendentry{AE-online}

\end{axis}

\end{tikzpicture}%
}\\
    \end{minipage} 
    \caption{Performance comparison under different compression dimensions when the observation dimension is 20.}
    \label{fig_Level1_ObsDim20_diffComp}
    \vspace{-6mm}
\end{figure*}

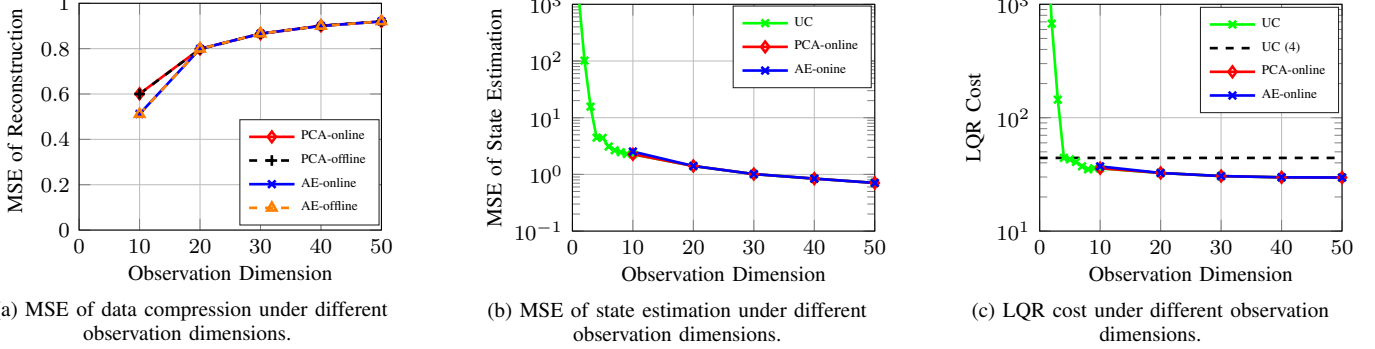
\begin{figure*}[tb]
    \centering
    \captionsetup[subfigure]{justification=centering, font=footnotesize, skip=5pt} 
    \begin{minipage}{0.32\textwidth}
    	\subfloat[MSE of data compression under different observation dimensions. \label{fig:MSE_CompDim4_diffObs}]{
%
\newcommand\ms{0.8}
\newcommand\lw{1.0}
\definecolor{mycolor1}{rgb}{0.00000,1.00000,1.00000}%
\begin{tikzpicture}[font=\footnotesize, spy using outlines={rectangle, magnification=2.5, size=0.3cm, connect spies}]
\begin{axis}[%
width=4.cm,
height=3.cm,
at={(0in,0in)},
scale only axis,
xmin=0,
xmax=50,
xtick={0,10,20,30,40,50},
xlabel style={yshift=0.5 ex},
xlabel={Observation Dimension},
ymin=0,
ymax=1,
ylabel style={xshift= 0.5 ex},
ylabel={MSE of Reconstruction},
axis background/.style={fill=white},
xmajorgrids,
ymajorgrids,
legend columns=1, 
legend style={font=\tiny, at={(0.53,0.02)}, anchor=south west, legend cell align=left, align=left, draw=white!15!black}
]

\addplot [color=red, line width=\lw pt, mark=diamond, mark options={solid, red}]
  table[row sep=crcr]{%
10	0.601\\
20	0.799\\
30	0.866\\
40	0.901\\
50	0.920\\
};
\addlegendentry{PCA-online}

\addplot [color=black, dashed, line width=\lw pt, mark=+, mark options={solid, black}]
  table[row sep=crcr]{%
10	0.600\\
20	0.798\\
30	0.866\\
40	0.901\\
50	0.920\\
};
\addlegendentry{PCA-offline}

\addplot [color=blue, line width=\lw pt, mark=x, mark options={solid, blue}]
  table[row sep=crcr]{%
10	0.512\\
20	0.799\\
30	0.866\\
40	0.900\\
50	0.920\\
};
\addlegendentry{AE-online}

\addplot [color=orange, dashed, line width=\lw pt, mark=triangle, mark options={solid, orange}]
  table[row sep=crcr]{%
10	0.511\\
20	0.799\\
30	0.866\\
40	0.900\\
50	0.920\\
};
\addlegendentry{AE-offline}


\end{axis}

\end{tikzpicture}%
}\\
    \end{minipage}
    \hfill
    \begin{minipage}{0.32\textwidth}
    	\subfloat[MSE of state estimation under different observation dimensions.\label{fig:state_CompDim4_diffObs}]{
%
\newcommand\ms{0.8}
\newcommand\lw{1.0}
\definecolor{mycolor1}{rgb}{0.00000,1.00000,1.00000}%
\begin{tikzpicture}[font=\footnotesize, spy using outlines={rectangle, magnification=2.5, size=0.3cm, connect spies}]
\begin{axis}[%
width=4.cm,
height=3.cm,
at={(0in,0in)},
scale only axis,
xmin=0,
xmax=50,
xtick={0,10,20,30,40,50},
xlabel style={yshift=0.5 ex},
xlabel={Observation Dimension},
ymode=log,
ymin=0.1,
ymax=1e3,
ytick={1e-1, 1e0, 1e1, 1e2, 1e3},
ylabel style={yshift=-0.5 ex},
ylabel={MSE of State Estimation},
axis background/.style={fill=white},
xmajorgrids,
ymajorgrids,
legend columns=1, 
legend style={font=\tiny, at={(0.53, 0.63)}, anchor=south west, legend cell align=left, align=left, draw=white!15!black}
]

\addplot [color=green, line width=\lw pt, mark=x, mark options={solid, green}]
  table[row sep=crcr]{%
1 1537  \\
2 101.8 \\
3 15.66 \\
4 4.450 \\
5 4.412  \\
6 3.101 \\
7 2.662   \\
8 2.470 \\
9 2.294 \\
10	2.264\\
20	1.407\\
30	1.013\\
40	0.842\\
50	0.707\\
};
\addlegendentry{UC}

\addplot [color=red, line width=\lw pt, mark=diamond, mark options={solid, red}]
  table[row sep=crcr]{%
10	2.265\\
20	1.407\\
30	1.013\\
40	0.842\\
50	0.707\\
};
\addlegendentry{PCA-online}

\addplot [color=blue, line width=\lw pt, mark=x, mark options={solid, blue}]
  table[row sep=crcr]{%
10	2.527\\
20	1.407\\
30	1.013\\
40	0.843\\
50	0.707\\
};
\addlegendentry{AE-onine}

\end{axis}

\end{tikzpicture}%
}\\
    \end{minipage}
    \hfill
    \begin{minipage}{0.32\textwidth}
    	\subfloat[LQR cost under different observation dimensions.\label{fig:LQR_CompDim4_diffObs}]{
%
\newcommand\ms{0.8}
\newcommand\lw{1.0}
\definecolor{mycolor1}{rgb}{0.00000,1.00000,1.00000}%
\begin{tikzpicture}[font=\footnotesize, spy using outlines={rectangle, magnification=2.5, size=0.3cm, connect spies}]
\begin{axis}[%
width=4.cm,
height=3.cm,
at={(0in,0in)},
scale only axis,
xmin=0,
xmax=50,
xtick={0,10,20,30,40,50},
xlabel style={yshift=0.5 ex},
xlabel={Observation Dimension},
ymode=log,
ymin=10,
ymax=1e3,
ylabel style={yshift=-0.5 ex},
ylabel={LQR Cost},
axis background/.style={fill=white},
xmajorgrids,
ymajorgrids,
legend columns=1, 
legend style={font=\tiny, at={(0.53, 0.52)}, anchor=south west, legend cell align=left, align=left, draw=white!15!black}
]

\addplot [color=green, line width=\lw pt, mark=x, mark options={solid, green}]
  table[row sep=crcr]{%
1 6221.5  \\
2 675.60\\
3 144.30\\
4 44.20\\
5 42.92\\
6 40.70 \\
7 37.26 \\
8 35.06\\
9 35.85\\
10	35.70\\
20	32.55\\
30	30.57\\
40	29.80\\
50	29.68\\
};
\addlegendentry{UC}

\addplot [color=black, line width=\lw pt, mark=., mark options={solid, black},dashed]
  table[row sep=crcr]{%
0 44.20\\
10 44.20 \\
20 44.20 \\
30 44.20 \\
40 44.20 \\
50 44.20 \\
};
\addlegendentry{UC (4)}

\addplot [color=red, line width=\lw pt, mark=diamond, mark options={solid, red}]
  table[row sep=crcr]{%
10	35.85 \\
20	32.56 \\
30	30.57 \\
40	29.80 \\
50  29.68 \\
};
\addlegendentry{PCA-online}

\addplot [color=blue, line width=\lw pt, mark=x, mark options={solid, blue}]
  table[row sep=crcr]{%
10	37.20 \\
20	32.56 \\
30	30.57 \\
40	29.81 \\
50  29.68 \\
};
\addlegendentry{AE-online}

\end{axis}

\end{tikzpicture}%
}\\
    \end{minipage}
    \caption{Performance comparison under different observation dimensions when the compression dimension is $\dim(\bm{z}_t^s)=4$ (equal to the state dimension).}
    \label{fig_Level1_CompDim4_diffObs}
   \vspace{-4mm}
\end{figure*}

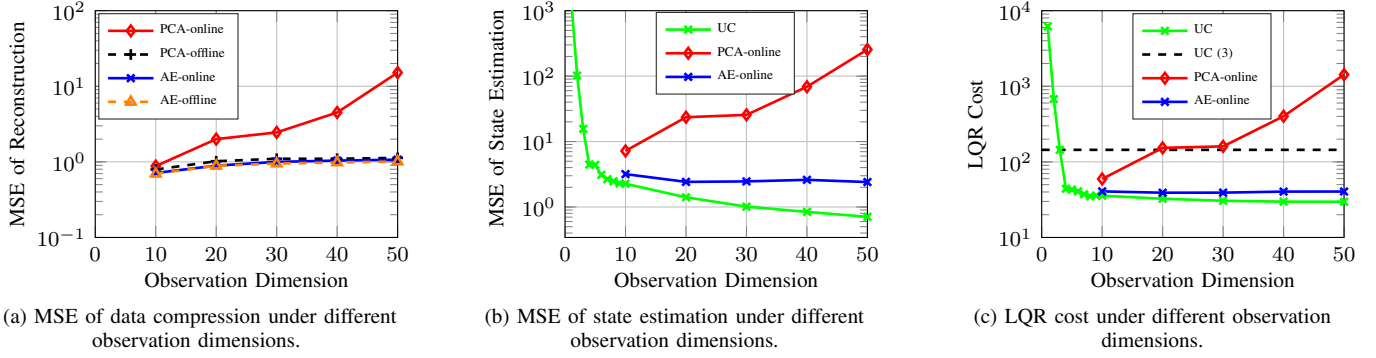
\begin{figure*}[t]
    \centering
    \captionsetup[subfigure]{justification=centering, font=footnotesize, skip=5pt} 
    \begin{minipage}{0.32\textwidth}    
    	\subfloat[MSE of data compression under different observation dimensions. \label{fig:MSE_CompDim3_diffObs}]{
%
\newcommand\ms{0.8}
\newcommand\lw{1.0}
\definecolor{mycolor1}{rgb}{0.00000,1.00000,1.00000}%
\begin{tikzpicture}[font=\footnotesize, spy using outlines={rectangle, magnification=2.5, size=0.3cm, connect spies}]
\begin{axis}[%
width=4.cm,
height=3.cm,
at={(0in,0in)},
scale only axis,
xmin=0,
xmax=50,
xtick={0,10,20,30,40,50},
xlabel style={yshift=0.5 ex},
xlabel={Observation Dimension},
ymode=log,
ymin=0.1,
ymax=100,
ylabel style={yshift=-0.5 ex},
ylabel={MSE of Reconstruction},
axis background/.style={fill=white},
xmajorgrids,
ymajorgrids,
legend columns=1, 
legend style={font=\tiny, at={(0.01,0.52)}, anchor=south west, legend cell align=left, align=left, draw=white!15!black}
]

\addplot [color=red, line width=\lw pt, mark=diamond, mark options={solid, red}]
  table[row sep=crcr]{%
10	0.879\\
20	2.001\\
30	2.454\\
40	4.5\\
50	15.173\\
};
\addlegendentry{PCA-online}

\addplot [color=black, dashed, line width=\lw pt, mark=+, mark options={solid, black}]
  table[row sep=crcr]{%
10	0.793\\
20	1.02\\
30	1.10\\
40	1.10\\
50	1.13\\
};
\addlegendentry{PCA-offline}

\addplot [color=blue, line width=\lw pt, mark=x, mark options={solid, blue}]
  table[row sep=crcr]{%
10	0.713\\
20	0.892\\
30	1.00\\
40	1.04\\
50	1.07\\
};
\addlegendentry{AE-online}

\addplot [color=orange, dashed, line width=\lw pt, mark=triangle, mark options={solid, orange}]
  table[row sep=crcr]{%
10	0.697\\
20	0.888\\
30	0.949\\
40	0.982\\
50	1.01\\
};
\addlegendentry{AE-offline}

\end{axis}

\end{tikzpicture}%
}\\
    \end{minipage}
    \hfill
    \begin{minipage}{0.32\textwidth}    
    	\subfloat[MSE of state estimation under different observation dimensions.\label{fig:state_CompDim3_diffObs}]{
%
\newcommand\ms{0.8}
\newcommand\lw{1.0}
\definecolor{mycolor1}{rgb}{0.00000,1.00000,1.00000}%
\begin{tikzpicture}[font=\footnotesize, spy using outlines={rectangle, magnification=2.5, size=0.3cm, connect spies}]
\begin{axis}[%
width=4.cm,
height=3.cm,
at={(0in,0in)},
scale only axis,
xmin=0,
xmax=50,
xtick={0,10,20,30,40,50},
xlabel style={yshift=0.5 ex},
xlabel={Observation Dimension},
ymode=log,
ymin=0,
ymax=1e3,
ylabel style={xshift=0.5 ex},
ylabel={MSE of State Estimation},
axis background/.style={fill=white},
xmajorgrids,
ymajorgrids,
legend columns=1, 
legend style={font=\tiny, at={(0.3, 0.63)}, anchor=south west, legend cell align=left, align=left, draw=white!15!black}
]

\addplot [color=green, line width=\lw pt, mark=x, mark options={solid, green}]
  table[row sep=crcr]{%
1 1537  \\
2 101.8 \\
3 15.66 \\
4 4.450 \\
5 4.412  \\
6 3.101 \\
7 2.662   \\
8 2.470 \\
9 2.294 \\
10	2.264\\
20	1.407\\
30	1.013\\
40	0.842\\
50	0.707\\
};
\addlegendentry{UC}

\addplot [color=red, line width=\lw pt, mark=diamond, mark options={solid, red}]
  table[row sep=crcr]{%
10	7.20 \\
20	23.56 \\
30	25.57 \\
40	68.81 \\
50  254.68 \\
};
\addlegendentry{PCA-online}

\addplot [color=blue, line width=\lw pt, mark=x, mark options={solid, blue}]
  table[row sep=crcr]{%
10	3.18\\
20	2.42\\
30	2.46\\
40	2.60\\
50	2.4062\\
};
\addlegendentry{AE-online}

\end{axis}

\end{tikzpicture}%
}\\
    \end{minipage}
    \hfill
    \begin{minipage}{0.32\textwidth}
    	\subfloat[LQR cost under different observation dimensions.\label{fig:LQR_CompDim3_diffObs}]{
%
\newcommand\ms{0.8}
\newcommand\lw{1.0}
\definecolor{mycolor1}{rgb}{0.00000,1.00000,1.00000}%
\begin{tikzpicture}[font=\footnotesize, spy using outlines={rectangle, magnification=2.5, size=0.3cm, connect spies}]
\begin{axis}[%
width=4.cm,
height=3.cm,
at={(0in,0in)},
scale only axis,
xmin=0,
xmax=50,
xtick={0,10,20,30,40,50},
xlabel style={yshift=0.5 ex},
xlabel={Observation Dimension},
ymode=log,
ymin=10,
ymax=1e4,
ylabel style={yshift=-0.5 ex},
ylabel={LQR Cost},
axis background/.style={fill=white},
xmajorgrids,
ymajorgrids,
legend columns=1, 
legend style={font=\tiny, at={(0.30,0.52)}, anchor=south west, legend cell align=left, align=left, draw=white!15!black}
]

\addplot [color=green, line width=\lw pt, mark=x, mark options={solid, green}]
  table[row sep=crcr]{%
1 6221.5  \\
2 675.60\\
3 144.30\\
4 44.20\\
5 42.92\\
6 40.70 \\
7 37.26 \\
8 35.06\\
9 35.85\\
10	35.70\\
20	32.55\\
30	30.57\\
40	29.80\\
50	29.68\\
};
\addlegendentry{UC}

\addplot [color=black, line width=\lw pt, mark=., mark options={solid, black}, dashed]
  table[row sep=crcr]{%
0 144.30\\
10 144.30\\
20 144.30\\
30 144.30\\
40 144.30\\
50 144.30\\
};
\addlegendentry{UC (3)}

\addplot [color=red, line width=\lw pt, mark=diamond, mark options={solid, red}]
  table[row sep=crcr]{%
10	59.43 \\
20	153.14 \\
30	160.39 \\
40	401.70 \\
50	1426.23 \\
};
\addlegendentry{PCA-online}

\addplot [color=blue, line width=\lw pt, mark=x, mark options={solid, blue}]
  table[row sep=crcr]{%
10	40.68 \\
20	39.08 \\
30	39.19 \\
40	40.43 \\
50  40.51 \\
};
\addlegendentry{AE-online}

\end{axis}

\end{tikzpicture}%
}\\
    \end{minipage}
    \caption{Performance comparison under different observation dimensions when the compression dimension is $\dim(\bm{z}_t^s)=3$ (smaller than the state dimension).}
    \label{fig_Level1_CompDim3_diffObs}
    \vspace{-5mm}
\end{figure*}

\vspace{-1mm}
\subsection{Results for A Single Sensor}

We evaluate system performance in a single-sensor setting, focusing on different observation and compression dimensions.

\subsubsection{Impact of Varying Compression Dimensions}
Firstly, we evaluate system performance under different compression dimensions, with the observation dimension $N^s_y$ fixed at 20 in Fig.~\ref{fig_Level1_ObsDim20_diffComp}.  As shown in Fig.~\ref{fig_Level1_ObsDim20_diffComp}\subref{fig:MSE_ObsDim20_diffComp}, the AE consistently achieves lower reconstruction error than PCA, demonstrating its stronger representation capacity. As shown in Fig.~\ref{fig_Level1_ObsDim20_diffComp}\subref{fig:state_ObsDim20_diffComp} and Fig.~\ref{fig_Level1_ObsDim20_diffComp}\subref{fig:LQR_ObsDim20_diffComp}, when the compression dimension exceeds 4, the LQR cost remains nearly identical to that of the non-compression results, i.e., UC (20). However, when the compression dimension falls below 4, the LQR cost increases significantly as the compression dimension decreases. This result leads to an important insight: Compression is lossless when the compression dimension is greater than or equal to the system's state dimension. Below this threshold, compression becomes lossy and degrades control performance. Additionally, comparison with the uncompressed case (UC), where the observation and compression dimensions are equal, shows that compression-based methods can outperform UC when the compression dimension exceeds 4.

\vspace{-1mm}
\subsubsection{Lossless Compression at the State Dimension}
We plot the results for the case where the compression dimension is fixed to the system’s state dimension, i.e., $\dim(\bm{z}_t^s)=N_x=4$ in Fig.~\ref{fig_Level1_CompDim4_diffObs}. As shown in Fig.~\ref{fig_Level1_CompDim4_diffObs}\subref{fig:MSE_CompDim4_diffObs}, the reconstruction error is nearly identical in both online and offline settings for PCA and AE. As the observation dimension increases, reconstruction error also increases due to a higher compression ratio, which leads to more information loss. However, as shown in Fig.~\ref{fig_Level1_CompDim4_diffObs}\subref{fig:state_CompDim4_diffObs} and Fig.~\ref{fig_Level1_CompDim4_diffObs}\subref{fig:LQR_CompDim4_diffObs}, both PCA and AE achieve state estimation and LQR performance comparable to the uncompressed baseline, indicating lossless at this compression level. Furthermore, both state estimation error and LQR cost decrease with higher observation dimensions, highlighting the benefit of richer sensing information.

\vspace{-1mm}
\subsubsection{Lossy Compression Below the State Dimension}
Fig.~\ref{fig_Level1_CompDim3_diffObs} shows the results when the compression dimension is reduced to 3 (below the state dimension, i.e., $\dim(\bm{z}_t^s)<N_x$). As shown in Fig.~\ref{fig_Level1_CompDim3_diffObs}\subref{fig:MSE_CompDim3_diffObs}, although PCA performs slightly worse than AE in offline reconstruction, this difference becomes critical in the online setting. AE’s lower reconstruction error helps maintain system stability, while PCA leads to instability and cumulative errors due to deviation from the LQR steady state. Simultaneously, as the observation dimension increases, the reconstruction errors of both PCA and AE also increase. For state estimation error and LQR cost, as shown in Fig.~\ref{fig_Level1_CompDim3_diffObs}\subref{fig:state_CompDim3_diffObs} and Fig.\ref{fig_Level1_CompDim3_diffObs}\subref{fig:LQR_CompDim3_diffObs}, PCA performance degrades as the observation dimension increases, due to a higher compression ratio that limits the amount of transmittable information. This results in larger reconstruction and state estimation errors. In contrast, AE performance remains stable, as it can better extract useful features from high-dimensional inputs.

\begin{figure}[tb]
    \centering
    \captionsetup[subfigure]{justification=centering, font=footnotesize, skip=3pt} 
    \hspace*{-3.3em}
    \begin{minipage}{0.23\textwidth}        
    	\subfloat[Reconstruction MSE for \\ fixed total transmission \\dimension.\label{fig_fixed_transmission_dimension_MSE}]{\newcommand\ms{0.8}
\newcommand\lw{1.0}
\definecolor{mycolor1}{rgb}{0.00000,1.00000,1.00000}

\begin{tikzpicture}[font=\footnotesize, spy using outlines={rectangle, magnification=2.5, size=0.3cm, connect spies}]
\begin{axis}[%
    width=2.8cm,
    height=2.8cm,
    at={(0in,0in)},
    scale only axis,
    xmin=1, xmax=5,
    xlabel style={xshift=-1ex,yshift=0.5ex},
    xlabel={Compression Dimension of Sensor 1},
    ymode=log,
    ymin=1e-1, ymax=1e3,
    ytick={1e-1, 1e0, 1e1, 1e2, 1e3},
    ylabel style={yshift=-1.5ex},
    ylabel={MSE of Reconstruction},
    axis background/.style={fill=white},
    xmajorgrids, ymajorgrids,
    legend columns=1, 
    legend style={font=\fontsize{5}{5.5}\selectfont, at={(0.465, 0.77)}, anchor=south west, legend cell align=left, align=left, fill opacity=0.7, draw opacity=1, text opacity=1, inner sep=1pt, inner xsep=1pt, row sep=-2pt, draw=white!15!black}
]

\addplot [color=blue, line width=\lw pt, mark=x, mark options={solid, blue}]
  table[row sep=crcr]{%
1 46949 \\
2 111 \\
3 0.88 \\
4 0.80 \\
5 0.694  \\
};
\addlegendentry{Sensor 1}

\addplot [color=red, line width=\lw pt, mark=diamond, mark options={solid, red}]
  table[row sep=crcr]{%
1 9732.73 \\
2 242.29\\
3 8.00 \\
4 9.22 \\
5 25.77 \\
};
\addlegendentry{Sensor 2}

\end{axis}
\end{tikzpicture}}\\
    \end{minipage}
    \hspace{-0.2em}
    \begin{minipage}{0.23\textwidth}    
    	\subfloat[State estimation error and \\ LQR cost under fixed total \\ transmission dimensions.\label{fig_fixed_transmission_dimension_LQR}]{\newcommand\ms{0.8}
\newcommand\lw{1.0}
\definecolor{mycolor1}{rgb}{0.00000,1.00000,1.00000}

\begin{tikzpicture}[font=\footnotesize, spy using outlines={rectangle, magnification=2.5, size=0.3cm, connect spies}]
\begin{axis}[%
    width=2.8cm,
    height=2.8cm,
    at={(0in,0in)},
    scale only axis,
    xmin=1, xmax=5,
    xlabel style={yshift=0.5ex,xshift=0.1ex},
    xlabel={Compression Dimension of Sensor 1},
    ymode=log,
    ymin=10, ymax=1e4,
    ylabel style={yshift=-1.5ex},
    ylabel={LQR Cost},
    axis background/.style={fill=white},
    xmajorgrids, ymajorgrids,
    legend columns=1, 
    legend style={font=\fontsize{5}{5.5}\selectfont, at={(0.26, 0.772)}, anchor=south west, legend cell align=left, align=left, fill opacity=0.7, draw opacity=1, text opacity=1, inner sep=1pt, inner xsep=1pt, row sep=-2pt, draw=white!15!black}
]
\addplot [color=blue, line width=\lw pt, mark=x, mark options={solid, blue}]
  table[row sep=crcr]{%
1 4068655 \\
2 8028 \\
3 37.26 \\
4 32.41 \\
5 42.2  \\
};

\end{axis}

\begin{axis}[%
    width=2.8cm,
    height=2.8cm,
    at={(0in,0in)},    
    scale only axis,
    xmin=1, xmax=5,
    ymin=0, ymax=10000,
    ymode=log,
    ylabel style={yshift=1.5ex},
    ylabel={MSE of State Estimation},
    axis y line*=right,
    axis x line=none,
    ymajorgrids=false,
    legend columns=1, 
    legend style={font=\fontsize{5}{5.5}\selectfont, at={(0.26, 0.772)}, anchor=south west, legend cell align=left, align=left, fill opacity=0.7, draw opacity=1, text opacity=1, inner sep=1pt, inner xsep=1pt, row sep=-2pt, draw=white!15!black}
]

\addlegendimage{mark=x, color=blue, line width=\lw pt}
\addlegendentry{LQR Cost}
\addplot [color=red, line width=\lw pt, mark=o, mark options={solid, red}]
  table[row sep=crcr]{%
1 948443 \\
2 1420 \\
3 2.2 \\
4 1.4 \\
5 3.9 \\
};
\addlegendentry{Estimation Error}

\end{axis}
\end{tikzpicture}}\\
    \end{minipage} 
    \caption{LQR cost when the total transmission dimension is fixed at 6. The x-axis represents the compression dimension of sensor 1, i.e., $\dim(\bm{z}_t^1)$, while the compression dimension of sensor 2 is given by $\dim(\bm{z}_t^2)= 6 - \dim(\bm{z}_t^1)$.}
    \label{fig_fixed_transmission_dimension}
    \vspace{-4mm}
\end{figure}
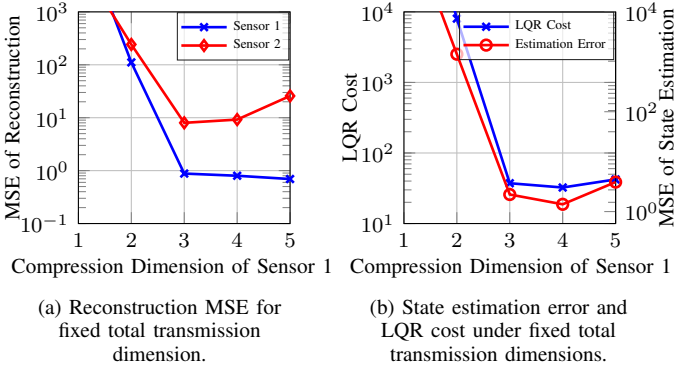

\vspace{-1mm}
\subsection{Results for Multiple Sensors}

\subsubsection{Performance with Fixed Transmission Dimension}
In the multi-sensor scenario, we consider two distributed sensors, each with an observation dimension of 20. Sensor 1 has low noise with $\bm{R}^1 = \bm{I}$, while sensor 2 has higher noise with $\bm{R}^2 = 10\bm{I}$. The resource allocation strategy is specified as $(\dim(\bm{z}_t^1), \dim(\bm{z}_t^2))$, denoting the compression dimensions assigned to sensors 1 and 2, respectively. We first evaluate performance under a fixed total transmission dimension $r_t$, comparing different allocation strategies. We then examine the optimal performance and corresponding allocations across varying $r_t$.

\begin{table}[tb]
    \centering
    \caption{Comparison of Compression and Non-Compression Results when the total transmission dimension is fixed to 6}
    \label{tab:compression_results}
    \begin{tabular}{l|c}
        \hline
        \textbf{Method} & \textbf{LQR Cost} \\
        \hline
        Optimal allocation strategy (4,2) & 32.41 \\
        UC (Sensor 1) & 32.56 \\
        UC (Sensor 2) & 54.44 \\
        UC (20) & 31.94 \\
        UC (3) & 113.52 \\
        \hline
    \end{tabular}
   \vspace{-4mm}
\end{table}

Fig.~\ref{fig_fixed_transmission_dimension} illustrates the system performance when the total transmission dimension is fixed at 6, i.e., $\dim(\bm{z}_t^1)+\dim(\bm{z}_t^2) = r_t =6$. As shown in Fig.~\ref{fig_fixed_transmission_dimension}\subref{fig_fixed_transmission_dimension_LQR}, when sensor 1's compression dimension increases, the LQR cost and state estimation error first decrease and then increase. The best result occurs at the allocation (4,2). As sensor 1 (the low-noise sensor) provides more reliable observations, increasing its compression dimension improves performance until lossless compression is effectively achieved at dimension 4. Beyond that, allocating additional resources to sensor 1 (e.g., (5,1)) reduces the transmission dimension available to sensor 2, discarding potentially valuable complementary information. Meanwhile, Fig.~\ref{fig_fixed_transmission_dimension}\subref{fig_fixed_transmission_dimension_MSE} shows the reconstruction error of sensors 1 and 2 across allocation strategies. As sensor 1's compression dimension increases, its reconstruction MSE decreases consistently. For sensor 2, as the compression dimension of sensor 1 increases, the MSE first decreases and then increases. This rise is due to two factors: initially, when the compression dimension of sensor 1 is low but that of sensor 2 is high, the LQR system deviates from its stable state, causing the sensor observations to fall outside the training distribution of AE and degrading the reconstruction quality. Later, as the compression dimension of sensor 1 increases, sensor 2 suffers from lossy compression at a lower compression dimension, and the reconstruction error increases again.

\begin{figure}[tb]
    \centering
    \captionsetup[subfigure]{justification=centering, font=footnotesize, skip=5pt} 
    \begin{minipage}{0.4\textwidth}        
    	\subfloat[MSE of reconstruction under different compression dimensions for sensor 1 and 2.\label{fig_total_transmission_dimension_MSE}]{\newcommand\ms{0.8}
\newcommand\lw{1.0}
\definecolor{mycolor1}{rgb}{0.00000,1.00000,1.00000}

\begin{tikzpicture}[font=\footnotesize, spy using outlines={rectangle, magnification=2.5, size=0.3cm, connect spies}]
\begin{axis}[%
    width=4.3cm,
    height=2.8cm,
    at={(0in,0in)},
    scale only axis,
    xmin=1, xmax=5,
    xlabel style={yshift=0.5ex},
    xlabel={Compression Dimension of Each Sensor},
    ymode=log,
    ymin=0.1, ymax=1e3,
    ytick={1e-1, 1e0, 1e1, 1e2, 1e3},
    ylabel style={yshift=-1.5ex},
    ylabel={MSE of Reconstruction},
    axis background/.style={fill=white},
    xmajorgrids, ymajorgrids,
    legend columns=1, 
    legend style={font=\tiny, at={(0.61, 0.71)}, anchor=south west, legend cell align=left, align=left, draw=white!15!black}
]

\addplot [color=blue, line width=\lw pt, mark=x, mark options={solid, blue}]
  table[row sep=crcr]{%
1  3e6 \\
2  117  \\
3  0.892 \\
4  0.800 \\
5  0.682 \\
};
\addlegendentry{Sensor 1}

\addplot [color=red, line width=\lw pt, mark=o, mark options={solid, red}]
  table[row sep=crcr]{%
1 6e6 \\
2 4e6\\
3 8.04 \\
4 6.75 \\
5 4.72 \\
};
\addlegendentry{Sensor 2}

\end{axis}
\end{tikzpicture}}\\
    \end{minipage}
    \hfill
    \begin{minipage}{0.4\textwidth}    
    	\subfloat[State estimation error and LQR cost under the optimal allocation determined by the LQR cost.\label{fig_total_transmission_dimension_LQR}]{\newcommand\ms{0.8}
\newcommand\lw{1.0}
\definecolor{mycolor1}{rgb}{0.00000,1.00000,1.00000}

\begin{tikzpicture}[font=\footnotesize, spy using outlines={rectangle, magnification=2.5, size=0.3cm, connect spies}]
\begin{axis}[%
    width=4.3cm,
    height=2.8cm,
    at={(0in,0in)},
    scale only axis,
    xmin=2, xmax=8,
    xlabel style={yshift=0.5ex},
    xlabel={Total Transmission Dimension},
    ymode=log,
    ymin=10, ymax=1e4,
    ylabel style={yshift=-1.5ex},
    ylabel={LQR Cost},
     ylabel style={blue},
    axis background/.style={fill=white},
    xmajorgrids, ymajorgrids,
    legend columns=1, 
    legend style={font=\tiny, at={(0.31, 0.495)}, anchor=south west, legend cell align=left, align=left, draw=white!15!black}
]

\addplot [color=blue, line width=\lw pt, mark=x, mark options={solid, blue}]
  table[row sep=crcr]{%
2 8547 \\
3 39.01 \\
4 32.56 \\
5 32.56 \\
6 32.41 \\
7 32.05 \\
8 32.02 \\
};
\addlegendentry{LQR Cost}

\addplot [color=blue, line width=\lw pt, mark=., mark options={solid, blue}, dashed]
  table[row sep=crcr]{%
2 31.94 \\
8 31.94 \\
};
\addlegendentry{UC (20), LQR Cost}

\addlegendimage{mark=o, color=red, line width=\lw pt}
\addlegendentry{Estimation Error}

\addlegendimage{mark=., color=red, line width=\lw pt, dashed}
\addlegendentry{UC (20), Estimation Error}

\node at (axis cs:2,0.65e4) [anchor=west, font=\scriptsize, blue] {(2,0)};
\node at (axis cs:2.9,   59 ) [anchor=west, font=\scriptsize, blue] {(3,0)};
\node at (axis cs:3.6,   48 ) [anchor=west, font=\scriptsize, blue] {(4,0)};
\node at (axis cs:4.5,   47 ) [anchor=west, font=\scriptsize, blue] {(5,0)};
\node at (axis cs:5.5,   46 ) [anchor=west, font=\scriptsize, blue] {(4,2)};
\node at (axis cs:6.4,   46 ) [anchor=west, font=\scriptsize, blue] {(4,3)};
\node at (axis cs:7.2,   46 ) [anchor=west, font=\scriptsize, blue] {(4,4)};

\end{axis}

\begin{axis}[%
    width=4.3cm,
    height=2.8cm,
    at={(0in,0in)},    
    scale only axis,
    xmin=2, xmax=8,
    ymin=0, ymax=10000,
    ymode=log,
    ylabel style={yshift=1.5ex},
    ylabel={MSE of State Estimation},
     ylabel style={red},
    axis y line*=right,
    axis x line=none,
    ymajorgrids=false,
    legend style={font=\tiny, at={(0.6, 0.74)}, anchor=south west}
]

\addplot [color=red, line width=\lw pt, mark=o, mark options={solid, red}]
  table[row sep=crcr]{%
2 1493 \\
3 2.43 \\
4 1.40 \\
5 1.41 \\
6 1.39 \\
7 1.35 \\
8 1.33 \\
};

\addplot [color=red, line width=\lw pt, mark=., mark options={dashed, red}, dashed]
  table[row sep=crcr]{%
2 1.32 \\
8 1.32 \\
};

\end{axis}
\end{tikzpicture}}\\
    \end{minipage} 
    \caption{{Performance of the multi-sensor system with varying resource allocation and transmission dimensions.}}
    \label{fig_total_transmission_dimension}
   \vspace{-4mm}
\end{figure}
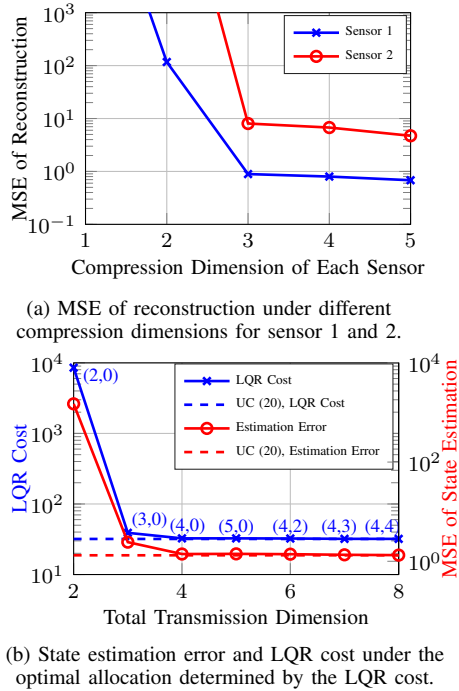

In Table \ref{tab:compression_results}, we compare the optimal allocation strategy that minimizes the LQR cost with several non-compressive baselines. The optimal strategy outperforms both single-sensor uncompressed schemes with an observation dimension of 20 (UC (Sensor 1) and UC (Sensor 2)). It shows that when the total transmission dimension exceeds the state dimension, leveraging observations from multiple sensors, even if partially lossy, can improve control performance. Furthermore, compared to UC (20) (where both sensors transmit uncompressed data, and each sensor has an observation dimension of 20), the optimal LQR cost from strategy (4,2) causes performance loss. This is because the compression of sensor 2 is lossy. However, compared to UC (3) (each sensor has an observation dimension of 3), the optimal strategy achieves a significant performance improvement.

\subsubsection{Performance under Different Total Transmission Dimension}

Fig.~\ref{fig_total_transmission_dimension} presents the reconstruction errors of the two sensors, as well as the optimal results and corresponding resource allocation strategies under different total transmission dimensions. In Fig.~\ref{fig_total_transmission_dimension}\subref{fig_total_transmission_dimension_MSE}, the reconstruction error of sensor 1 is consistently lower than that of sensor 2. As shown in Fig.~\ref{fig_total_transmission_dimension}\subref{fig_total_transmission_dimension_LQR}, the LQR cost decreases as the total transmission dimension increases. Regarding the optimal resource allocation, the strategy tends to assign more bandwidth to the sensor with lower observation noise (Sensor 1) as the total transmission budget increases. However, once sensor 1’s compression dimension exceeds 4, compression becomes effectively lossless. At this point, the strategy shifts to allocate additional resources to sensor 2, despite its higher observation noise, in order to further improve overall system performance.

\section{Conclusion and Future Work}
In this paper, we propose a general framework for closed-loop distributed ISAC systems. To address the problem of transmitting high-dimensional observations over rate-limited links, we formulate the observation compression problem and introduce an AE-based method to learn nonlinear representations. Through a detailed case study based on the LQR control scenario, we explore the interplay among state, observation, and compression dimensions, highlighting how compression impacts reconstruction accuracy, state estimation, and overall control performance. Additionally, we investigate optimal transmission resource allocation strategies for distributed multi-sensor systems under different sensor noise conditions. Although this work primarily focuses on how reconstruction errors impact the performance of closed-loop ISAC systems, future research will explore how semantic information can be further leveraged to improve compression efficiency in rate-limited, closed-loop distributed ISAC systems beyond LQR (i.e., with more complex non-linear sensing models, more complex state trackers, such as the extended Kalman filter, and more complex control policies, such as model-predictive control). In addition,  the work should be extended 
across the three layers of reconstruction, estimation, and control errors. 
\vspace{-1mm}
\balance 
\bibliographystyle{IEEEtran}
\bibliography{IEEEabrv, ref}

\end{document}